\documentclass
[aps,prb,twocolumn,showpacs,preprintnumbers,amsmath,amssymb,superscriptaddress]{revtex4}

\usepackage{graphicx}
\usepackage{dcolumn}
\usepackage{amsmath,amssymb}

\newcommand{\bm}[1]{\mbox{\boldmath{$#1$}}}

\def\ra{\rangle}
\def\la{\langle}
\def\dag{^\dagger}
\def\om{\Omega}

\def\cos{{\rm cos}}
\def\sin{{\rm sin}}

\makeatletter
\def\btt#1{\texttt{\@backslashchar#1}}%
\DeclareRobustCommand\bblash{\btt{\@backslashchar}}\makeatother

\begin{document}

\title[Short Title]{Topological Classification of Gapped Spin Chains
:Quantized Berry Phase as a Local Order Parameter}
\author{T. Hirano}
\email{hirano@pothos.t.u-tokyo.ac.jp}
\affiliation{%
Department of Applied Physics, The University of Tokyo, 7-3-1 Hongo, Bunkyo-ku, Tokyo 113-8656, Japan }%
\author{H. Katsura}%
\email{katsura@appi.t.u-tokyo.ac.jp}
\affiliation{%
Department of Applied Physics, The University of Tokyo, 7-3-1 Hongo, Bunkyo-ku, Tokyo 113-8656, Japan }%
\author{Y. Hatsugai}%
\email{hatsugai@sakura.cc.tsukuba.ac.jp}
\affiliation{%
Department of Applied Physics, The University of Tokyo, 7-3-1 Hongo, Bunkyo-ku, Tokyo 113-8656, Japan }%
\affiliation{%
Institute of Physics, 
University of Tsukuba, 1-1-1 Tennodai, Tsukuba, Ibaraki 305-8571, Japan }%
\date{\today}
\begin{abstract}
We characterize several phases 
of gapped spin systems by
 local order parameters defined by quantized Berry
 phases\cite{HatsugaiOrder3}. 
This characterization is
 topologically stable against any small perturbation
as long as the energy gap remains finite.
The models we pick up are 
 $S=1,2$ dimerized Heisenberg chains 
and  $S=2$ Heisenberg chains with uniaxial single-ion-type anisotropy. 
Analytically we also evaluate the topological local order parameters for the 
generalized Affleck-Kennedy-Lieb-Tasaki (AKLT) model.
The relation between the present Berry phases and the fractionalization in the
integer spin chains are discussed as well.
\end{abstract}

\maketitle
\section{Introduction}
Characterizing quantum many-body systems 
is one of the important topics
 in condensed matter physics.
The Ginzburg-Landau (GL) theory has been quite successful
to describe many phases based on a concept of 
the symmetry breaking and the local order parameter.
Despite its remarkable success, novel types of phases
which are not  well described by the (classical)
 local order parameters
have been found in many systems.
Concepts of 
topological order and quantum order are trial to overcome the
difficulties of the classical GL theory with the symmetry breaking
\cite{TOrder,HatsugaiOrder1,HatsugaiOrder2}. 
One of the characteristic features of
the  topological insulators is  that 
localized states, 
 such as the edge
states, appear  near the system boundaries,
even though the system without boundaries
has a finite energy gap.
Examples of such systems are 
quantum Hall liquids\cite{LaughlinEdge,HatsugaiEdge,HatsugaiER}, 
Haldane spin systems\cite{KennedyEdge,Hagiwara},
polyacetylene\cite{SSH}, and spin-Peierls system\cite{ReadSachdev}.
Recently it has become clear that the  bulk-edge correspondence 
\cite{HatsugaiEdge,HatsugaiER}
has an intimate relation to the entanglement entropy
\cite{Ryu,Hirano1,Katsura},
which has been discussed to detect non trivial structures
of topologically ordered states
\cite{TEE1,TEE2,Hirano1,Katsura}.

Recently one of the authors proposed to use another quantum quantity,
quantized Berry phases\cite{Berryphase}, to define a topological local order
parameter
\cite{HatsugaiOrder1,HatsugaiOrder2,HatsugaiOrder3}. 
One can define a topological local order parameter
by the Berry phases even though there is no classical order parameter.
The Berry phases is a typical quantum quantity based on the Berry connection
which is defined by the overlap between the two states with
infinitesimal difference.
It implies that the topological local order parameter defined in
\cite{HatsugaiOrder3}
is a quantum order parameter that does not have 
any corresponding classical analogs.
Further it has a conceptual advantage
for the topologically ordered
phases, since it is 
quantized to $0$ or $\pi$ (mod $2\pi$)
 when 
the ground state is invariant under some anti-unitary transformation.
It implies a topological stability that the quantized Berry phase does
not change
against any small perturbation.
The Berry phase is given by an integration of the 
Berry connection defined by the local $U(1)$ twist 
on a link of a lattice.
Then  the quantum phases can be categorized 
by the texture pattern of the Berry phases(0 or $\pi$). It  has been
successfully applied to several gapped quantum systems.
For example, the ground states of 
the $S=1/2$ dimerized Heisenberg models (in one and two dimensions
even with frustrations)\cite{HatsugaiOrder3}
can be characterized by the pattern of $\pi$ Berry phases on the 
bonds 
 which indicate  the locations of dimer singlets. 
In a case of 
the $t$-$J$ model\cite{Maruyama}, it is characterized by
the texture pattern of the non-Abelian Berry phase,
which describes
 itinerant singlets.
Also, for the $S=1$ Heisenberg model, its 
ground state as the Haldane phase
was characterized by the uniform $\pi$ Berry phases.
This topological order parameter also 
clearly describe a quantum phase transition between the
Haldane phase and the large-D phases\cite{HatsugaiOrderP}.

In this paper, we calculate the topological local order
parameter by the quantized Berry phase
for several gapped quantum spin chains.
There are substantial numbers of studies for
the Haldane phase
\cite{HaldaneGap,NENP,MnCl3,thetaterm}.
Then it has 
been clarified that the Haldane phase can be characterized by the hidden
$\mathbb{Z}_2\times\mathbb{Z}_2$
symmetry breaking\cite{Z2Z2-1,Z2Z2-2}
which describes a non-locality of the 
Haldane phase
(by the string order parameters) \cite{Nijs,Todo,Todo2}.
On the other hand, the topological order parameter by the Berry phases
is local and  
 quite useful for the $S=1$ case to describe the phase and the quantum phase 
transition\cite{HatsugaiOrderP}. 
Here we further investigate generic situations, such as 
the several Haldane phases in the dimerized $S=1,2$ 
Heisenberg chains\cite{Hida,OshikawaDimer} 
and the $S=2$ Heisenberg chain with uniaxial single-ion-type 
anisotropy\cite{HaldaneLargeD,SchollwockDterm,OshikawaDterm,OshikawaDterm2}. 
We also study the Berry phase of 
the generalized valence-bond-solid (VBS) state analytically and interpret
the numerical results in terms of the reconstruction of the
valence-bonds.

\section{Definition of the Berry phase}
Let us start with defining the Berry phase in a quantum spin system.
The Berry phase is defined when the Hamiltonian has parameters
with periodicity assuming a finite energy  gap 
between the ground state and the
excited states\cite{Berryphase}.
For the parameter dependent Hamiltonian $H(\phi)$, the
Berry phase $\gamma$ of the ground state is 
defined as 
\begin{eqnarray}
i\gamma&=&\int_0^{2\pi}A(\phi)d\phi,
\end{eqnarray}
 where $A(\phi)$ is the
Abelian Berry connection obtained 
by the single-valued
normalized ground state $|\mbox{GS}(\phi)\rangle$ of $H(\phi)$ as 
$A(\phi)=\langle\mbox{GS}(\phi)|\partial_\phi|\mbox{GS}(\phi)\rangle$.
This Berry phase is real and quantized to $0$ or $\pi$
(mod $2\pi$) if the Hamiltonian $H(\phi)$ is invariant
 under the anti-unitary operation $\Theta$, {\it i.e.} $[H(\phi),\Theta]=0$
\cite{HatsugaiOrder1}.
Note that the Berry phase is ``undefined''
if the gap between the ground state and 
the excited states vanishes while varying the parameter $\phi$.
We use a local spin twist on a link 
as a generic parameter in the definition of the Berry phase
\cite{HatsugaiOrder3}. 
Under this local spin twist, the following term 
$S_i^{+}S_j^{-}+S_i^{-}S_j^{+}$
in the Hamiltonian is replaced with
$e^{i\phi}S_i^{+}S_j^{-}+e^{-i\phi}S_i^{-}S_j^{+}$, 
where $S^{\pm}_i=S_i^x\pm iS_i^y$.
The Berry phase defined by the response to the
local spin twists extracts a local structure of the quantum system. 
By this  quantized Berry phase, one can define a link-variable.
Then  
each link has one of the three labels: ``$0$ bond'', ``$\pi$ bond'', or
``undefined''. 
It has a remarkable property that the Berry phase has topological
robustness against the small perturbations unless the
energy gap between the ground state and the excited states closes.
On the other hand, the ``undefined'' indicates an existence of the
quantum phase transition.
In order to calculate the Berry phase numerically,
we introduce a gauge-invariant Berry phase\cite{HatsugaiOrder3,Fukui}
on a lattice.
It is defined  by discretizing the parameter space of $\phi$
into $N$ points as 
\begin{eqnarray}
\gamma_N=-\sum_{n=1}^{N}\mbox{arg}A_N(\phi_n)\mbox{,}\ \phi_n=\frac {2\pi}Nn,
\end{eqnarray}
where $A_N(\phi_n)$ is defined by
$A_N(\phi_n)=\langle \mbox{GS}(\phi_n)|\mbox{GS}(\phi_{n+1})\rangle$,
$\phi_{N+1}=\phi_1$. We expect
$\gamma=\lim_{N\rightarrow\infty}\gamma_N$.
To calculate $\gamma_N$, we use the Lanczos method to diagonalize the
Hamiltonian in the subspace of $\sum_{i}S_i^{z}=0$.

\section{$S=1,2$ dimerized Heisenberg models and $S=2$ Heisenberg model
with uniaxial isotropy}
\subsection{$S=1,2$ dimerized Heisenberg models}
First  we consider  $S=1$, $2$ dimerized Heisenberg
models
\begin{eqnarray}
H&=&\sum_{i=1}^{N/2}
\left(
J_{1}\bm{S}_{2i}\cdot\bm{S}_{2i+1}+
J_{2}\bm{S}_{2i+1}\cdot\bm{S}_{2i+2}
\right),
\end{eqnarray}
where $\bm{S}_{i}$ is the spin-$1$ or $2$ operators on the $i$-th site and $N$ is the total number of sites.
The periodic boundary condition is imposed as $\bm{S}_{N+i}=\bm{S}_{i}$
for all of the models in this paper.
$J_1$ and $J_2$ are parameterized as $J_1=\sin{\theta}$ and
$J_2=\cos{\theta}$, respectively.
We consider the case of $0<\theta<\pi/2$ in this paper.
The ground state is composed of an ensemble of $N/2$ singlet pairs 
in limits of $\theta\to 0$ and $\theta\to\pi/2$.
The system is equivalent to the isotropic anti-ferromagnetic Heisenberg
chain at $\theta=\pi/4$.
Based on the VBS picture,
we expect a
reconstruction of the valence bonds by changing $\theta$.

\begin{figure}[!tb]
\includegraphics[width=7.5cm]{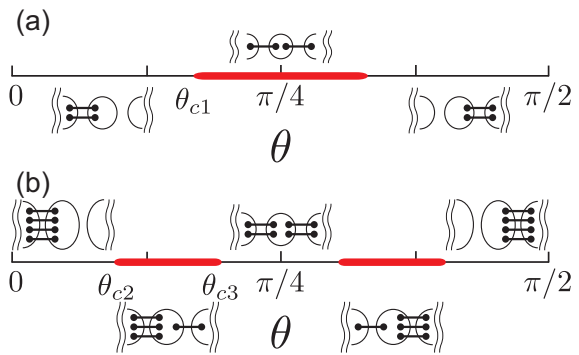}
\caption{(Color Online)
The Berry phases on the local link of (a) the $S=1$ periodic $N=14$ and 
(b) the $S=2$ periodic $N=10$ dimerized Heisenberg chains.
The Berry phase is $\pi$ on the bold line while that is 0 on the other line. 
The phase boundaries in the finite size system 
are $\theta_{c1}=0.531237$, $\theta_{c2}= 0.287453$ and $\theta_{c3}= 0.609305$, respectively. 
The Berry phase in (a) and (b) has an inversion symmetry with respect to
 $\theta=\pi/4$.
A schematic VBS picture of the ground state is assigned to each phase.
Dots, bold lines, and open circles denote the $S=1/2$, singlet dimers, 
and the operations of symmetrization, respectively.
}
\label{fig:Berry}
\end{figure}

\begin{figure}[!tb]
 \includegraphics[width=7.5cm]{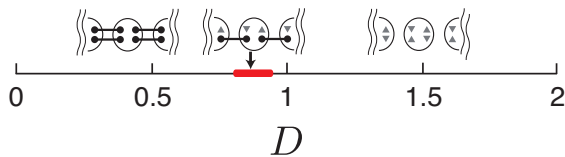}
\caption{(Color Online)
The Berry phases on the local link of
the $S=2$ periodic $N=10$  Heisenberg chain with single-ion anisotropy. 
The notations are the same as the Fig.\ref{fig:Berry}.
An up (down) triangle denotes an up (down) spin-$1/2$.
}
\label{fig:BerryD}
\end{figure}

Figures. \ref{fig:Berry}(a) and (b) show the $\theta$ dependence of the
Berry phase on the link with $J_1$ coupling and $J_2$ coupling with
$S=1$, $N=14$ and $S=2$, $N=10$, respectively.  The region with the
Berry phase $\pi$ is shown by the bold line.  
There are several quantum phase transitions characterized by
the Berry phase as the topological order parameters.
 The boundary of the two
regions with different Berry phases $0$ and $\pi$ does not have a well-defined
Berry phase, since the energy gap closes during the change of the local twist
parameter $\phi$.  
Since the Berry phase is
undefined at the boundaries, there exists the level crossing which
implies the existence of the gapless excitation in the thermodynamic
limit.  This result is consistent with 
the results previously discussed\cite{OshikawaDimer}, 
which the general integer-$S$ extended
string order parameters changes as the dimerization changes.
The phase diagram defined by our topological order parameter
is consistent with the one by the non-local string order parameter.
In an $N=10 $ system with $S=2$,
 the phase boundaries are $\theta_{c2}=
0.287453$, $\theta_{c3}= 0.609305$,
and it is consistent with the results obtained
by using the level spectroscopy which is based on conformal field
theory techniques\cite{Kitazawa}. 
Especially in the one dimensional case, the energy diagram of the 
system with twisted link is
proportional to that of the system with twisted boundary conditions.
However, our analysis focuses on the quantum property of the wave
functions rather than the energy diagram.

\subsection{$S=2$ Heisenberg model with uniaxial anisotropy}
As for the $S=2$
Heisenberg model with $D$-term,
we use the Hamiltonian 
\begin{eqnarray}
H &=& \sum_i^{N}
\left[
J{\bm S}_{i}\cdot{\bm S}_{i+1}+D\left(S_i^{z}\right)^2
\right].
\end{eqnarray}
Figure. \ref{fig:BerryD} shows the Berry phase of the
local link in the $S=2$ Heisenberg model + $D$-term
 with $N$=10. 
The parameter $J=1$ in our calculations.
The region of the bold line has the Berry phase $\pi$
and the other region has the vanishing Berry phase.
This result also makes us possible to consider the Berry phase as a local order
parameter of the Haldane spin chains.
Our numerical results for finite size systems 
support the presence of the intermediate $D$-phase 
\cite{OshikawaDterm}.

\section{Interpretation of the numerical results by valence-bond solid state}
Let us now interpret our numerical results in terms of the VBS state picture. 
The VBS state is the exact ground state of 
the Affleck-Kennedy-Lieb-Tasaki(AKLT) model\cite{AKLT}. 
We shall calculate the Berry phase of 
the generalized VBS state with the aid of the chiral AKLT model\cite{chiralVBS}
 and its exact ground state wave function. 
 The chiral AKLT model is obtained by applying O(2) rotation of spin operators in the original AKLT model. 
In our calculation, it is convenient to introduce the Schwinger boson representation of 
the spin operators 
as $S_i^{+}=a_i^{\dagger}b_{i}$, $S_i^{-}=a_ib_{i}^{\dagger}$, 
and $S_i^z=(a_i^{\dagger}a_{i}-b_i^{\dagger}b_{i})/2$.
$a_i$ and $b_i$ satisfy the commutation relation $[
a_i,a_j^{\dagger}]=[ b_i,b_j^{\dagger}]=\delta_{ij}$
with all other commutators vanishing \cite{Auelbach}.
The constraint $a_i^{\dagger}a_i+b_i^{\dagger}b_i=2S_i$ is imposed to reproduce
the dimension of the spin $S_i$ Hilbert space at each site.
In general, the ground state of the chiral AKLT model
having $B_{ij}$ valence bonds on the link $\la ij\ra$ is written as 
\begin{eqnarray}
|\{\phi_{i,j}\}\rangle&=&
\prod_{\langle ij\rangle}
\left(
e^{i\phi_{ij}/2}a^{\dagger}_{i}b^{\dagger}_{j}-e^{-i\phi_{ij}/2}b^{\dagger}_{i}a^{\dagger}_{j}
\right)^{B_{ij}}
|\mbox{vac}\rangle,
\end{eqnarray}
\cite{chiralVBS}.
This state has nonzero average of vector spin chirality $\la \bm{S}_i\times\bm{S}_j \cdot {\hat z}\ra$ unless 
the twist parameter $\phi_{ij}=0$ or $\pi$. 
This state is a zero-energy ground state of the following
Hamiltonian:
\begin{equation}
H(\{\phi_{i,i+1}\})=\sum_{i=1}^{N}\sum_{J=J_{\rm min}}^{J_{\rm max}}A_{J}
P^{J}_{i,i+1}[\phi_{i,i+1}],
\label{chAKLT}
\end{equation}
where $J_{\rm max}=\left(B_{i-1,i}+2B_{i,i+1}+B_{i+1,i+2}\right)/2$,
$J_{\rm min}=\left(B_{i-1,i}+B_{i+1,i+2}\right)/2+1$ and $A_J$ is the arbitrary positive coefficient.
$P^{J}_{i,i+1}[0]$ is the polynomial in
$\bm{S}_i\cdot\bm{S}_{i+1}$ and act as a projection
operator projecting the
bond spin $\bm{J}_{i,i+1}=\bm{S}_{i}+\bm{S}_{i+1}$ onto the subspace of 
spin magnitude $J$. The replacement
\begin{equation} 
S_i^+S_{i+1}^-+S_i^-S_{i+1}^+ \to e^{i\phi_{i,i+1}}S_i^+S_{i+1}^- + e^{-i \phi_{i,i+1}}S_i^-S_{i+1}^+ 
\end{equation}
in $\bm{S}_i\cdot\bm{S}_{i+1}$ 
produces $P^{J}_{i,i+1}[\phi_{i,i+1}]$ in Eq. (\ref{chAKLT}).

Now we shall explicitly show that the Berry phase of the VBS 
state extracts the local number of the valence bonds $B_{ij}$ as 
$B_{ij}\pi ({\rm mod}$ $2\pi$). Let us now consider the local twist of the parameters $\phi_{ij}=\phi \delta_{ij,12}$ and rewrite the ground state $|\{\phi_{i,j}\}\ra$ as $|\phi \ra$.
To calculate the Berry phase of the VBS state, the following relation is
useful: 
\begin{equation}
i\gamma_{12}=iB_{12}\pi+i
\int_0^{2\pi}{\rm Im}[\la \phi|\partial_\phi| \phi \ra ]/{\cal N}(\phi)d\phi,
\label{BPrelation}
\end{equation}
where $\gamma_{12}$ is the Berry phase of the bond $\la 12\ra$ and ${{\cal
N}(\phi)}=\la \phi|\phi \ra$. 
Note that the first term of the right hand side comes from the gauge fixing of the multi-valued wave function to the single-valued function. 
Then, the only thing to do is to evaluate the imaginary part of the
connection  

Let us first consider the $S=1$ VBS state as the simplest example. In this case, $B_{i,i+1}=1$ for any bond and the VBS state with a local twist is given by
\begin{equation}
|\phi\ra = \left(e^{i\phi/2}a^{\dagger}_1b^{\dagger}_2-e^{-i\phi/2}b^{\dagger}_1a^{\dagger}_2
\right) \prod_{i=2}^N
\left(a^{\dagger}_{i}b^{\dagger}_{i+1}-b^{\dagger}_{i}a^{\dagger}_{i+1}
\right)
|\mbox{vac}\rangle.
\end{equation}
We impose the periodic boundary condition, {\it i.e.}, $a_{N+1}=a_1$ and $b_{N+1}=b_1$. 
It is convenient to introduce the singlet creation operator $s^{\dagger}=(a_1^\dagger b_2^\dagger-b_1^\dagger a_2^\dagger)$ and the triplet
($J_z=0$) creation operator $t^{\dagger}=(a_1^\dagger b_2^\dagger+b_1^\dagger a_2^\dagger)$.
We can rewrite the bond $\la 12\ra$ part of the VBS state
$(e^{i\phi/2}a^{\dagger}_{1}b^{\dagger}_{2}-e^{-i\phi/2}b^{\dagger}_{1}a^{\dagger}_{2})$
as $(\cos{\frac{\phi}{2}}s^\dagger+i\sin{\frac{\phi}{2}}t^\dagger)$.
Then $|\phi\ra$  and $\partial_\phi|\phi\ra$ can be written as 
\begin{eqnarray}
|\phi \ra=\cos \frac{\phi}{2} |0\ra +i\sin \frac{\phi}{2} |1\ra,
\nonumber \\
\partial_\phi|\phi \ra = -\frac{1}{2}\sin \frac{\phi}{2} |0\ra +\frac{i}{2}\cos \frac{\phi}{2} |1\ra,
\end{eqnarray} 
where
\begin{eqnarray}
|0\ra &=&
s\dag\prod_{i=2}^N (a^{\dagger}_{i}b^{\dagger}_{i+1}-b^{\dagger}_{i}a^{\dagger}_{i+1})|{\rm vac}\ra,
\nonumber \\
|1\ra &=&
t\dag \prod_{i=2}^N(a^{\dagger}_{i}b^{\dagger}_{i+1}-b^{\dagger}_{i}a^{\dagger}_{i+1})|{\rm vac}\ra.
\nonumber
\end{eqnarray} 
It is now obvious that the imaginary part of $\la\phi|\partial_{\phi}|\phi\ra$ vanishes since the state $|1\ra$ having a total spin $S_{\rm total}=1$ is orthogonal to the state $|0\ra$ with $S_{\rm total}=0$. 
Therefore, the Berry phase of this state is given by $\gamma_{12}=\pi$. 
Next we shall consider a more general situation with arbitrary
$B_{ij}$. We can also express the VBS state with a local twist on the
bond $\la 12\ra$ in terms of $s\dag$ and $t\dag$ as
\begin{equation}
|\phi\ra = \Big(\cos{\frac{\phi}{2}}s^\dagger+i\sin{\frac{\phi}{2}}t^\dagger \Big)^{B_{12}}
\prod_{i=2}^N (a^{\dagger}_{i}b^{\dagger}_{i+1}-b^{\dagger}_{i}a^{\dagger}_{i+1})^{B_{i,i+1}}|{\rm vac}\ra.
\end{equation}
By using the binomial expansion, $|\phi\ra$ can be rewritten as 
\begin{equation}
|\phi\ra=\sum_{k=0}^{B_{12}} \binom {B_{12}} k \Big(\cos \frac{\phi}{2} \Big)^{B_{12}-k} \Big(i\sin \frac{\phi}{2} \Big)^k|k\ra,
\end{equation} 
where $|k\ra = (s\dag)^{B_{12}-k}(t\dag)^k(\cdots)|{\rm vac}\ra$ is the
state with $k$ triplet bonds on the link $\la 12\ra$. ($\cdots$) denotes the rest of the VBS state. 
In a parallel way, 
\begin{eqnarray}
\partial_{\phi}|\phi\ra &=& \frac{1}{2}\sum_{k=0}^{B_{12}} \binom{B_{12}}k \Big(\cos \frac{\phi}{2} \Big)^{B_{12}-k} \nonumber \\
&\times& \Big(i\sin \frac{\phi}{2} \Big)^k \Big(k\cot \frac{\phi}{2}-(B_{12}-k)\tan \frac{\phi}{2} \Big)|k\ra. \nonumber
\end{eqnarray} 
To see that the imaginary part of $\la\phi|\partial_{\phi}|\phi\ra$ is zero, we have to show that ${\rm Im}\la k|l\ra=0$ when $k$ and $l$ have the same parity(even or odd) and ${\rm Re}\la k|l \ra=0$ when $k$ and $l$ have different parities.  
This can be easily shown by using the coherent state representation of the Schwinger bosons (see APPENDIX). Then using the relation (\ref{BPrelation}),
 we can obtain the Berry phase 
as 
\[
\gamma_{ij}=B_{ij}\pi,\ \ \  (\text{mod } 2\pi).
\]
This result means that the Berry phase of the 
generalized VBS state counts
the number of the valence-bonds on the bond $\la ij\ra$.
One valence-bond has the $\pi$ Berry phase.
Finally, it should be stressed that 
our calculation of the Berry phase is not restricted to one-dimensional
VBS states but can be generalized to the VBS state on an {\it arbitrary graph}\cite{arbitVBS}
as long as there is a gap while varying the twist parameter.

Now, let us  consider 
the previous two models in terms of the VBS
picture. 
For the $S=2$ dimerized Heisenberg model, 
the number of the valence bonds changes as
the $\theta$ changes (see Fig.\ref{fig:Berry}).
Since the number of the valence bonds on a local link can be computed by
the Berry phase, we can clearly see that the reconstruction of the 
valence bonds occurs during the change of the dimerization. 
Thus, the result of the Berry phase is consistent with the VBS picture.
For the $S=2$ Heisenberg chain with single-ion anisotropy,
the valence bonds are broken one by one as $D$ increases
as we can see in the Fig. \ref{fig:BerryD}.
We see that the Berry phase reflects the number of the valence bonds
as well as the previous dimerized Heisenberg chain.
This can be understood as a fractionalization since the basic objects of
the present integer spin chains are spin-$1/2$ singlets.

\section{The relation between the Berry phase and the entanglement entropy}
Moreover, the Berry phase of generalized VBS state 
relates to the number of the edge states
which emerge when the spin chain has edges\cite{KennedyEdge}.
Thus, it detects the property of the topological phase.
Since the entanglement entropy also detects such
phases\cite{Katsura,Hirano1,Fan1},
we clarify the relation between the Berry phase and the entanglement entropy.
The entanglement entropy of our generalized VBS
state in thermodynamic limit is 
${\cal S}_A=\sum_{\la ij\ra\in \partial A}\log{(B_{ij}+1)}$,
where $\partial A$ denotes the set of the bonds on the boundary of
subsystem $A$.
It counts the number of the edge states $g_{\rm edge}$ as
$S_{A}=\log{g_{\rm edge}}$.
Thus, the Berry phase is related to the entanglement
entropy in generalized VBS states 
via the edge states in the thermodynamic limit\cite{Ryu}.

\section{Conclusion}
In conclusion, we have shown that the topological local order parameter
defined by quantized Berry phases is useful to classify the phases of
various spin chains such as the Haldane phase. 
In our calculations, 
the Berry phase is locally defined and does not need nonlocal calculations. 
It is also useful to estimate the order parameter from
the finite size systems since it is quantized even in the finite size systems.
The property of the phase is revealed in terms of the texture pattern of
the Berry phase. 
We have also analytically studied the Berry phase of the generalized VBS
state and found that the Berry phase picks up the number of singlets on
the local link. 
\begin{acknowledgments}
The authors are grateful to I. Maruyama and S. Todo for fruitful discussions.
The numerical diagonalization has been accomplished by utilizing 
the program package TITPACK ver. 2. The computation in this work has been
 done using the facilities of the Supercomputer Center, Institute for Solid State Physics, University of Tokyo.
HK was supported by the Japan Society for the Promotion of Science. 
YH was supported by Grants-in-Aid for Scientific Research
on Priority Areas from MEXT (No.18043007).
\end{acknowledgments}
\begin{appendix}
\section*{APPENDIX}
In this appendix we show that ${\rm Re} \la k |l \ra =0$ when $k$ and $l$ have different parities and ${\rm Im} \la k| l\ra=0$ when $k$ and $l$ have the same parity by simple symmetry arguments. To show them, it is convenient to introduce a spin coherent state \cite{Auelbach}. For a point ${\hat \Omega}=(\sin \theta \cos \phi, \sin \theta \sin \phi, \cos \theta)$ on the unit sphere, the spin coherent state at each site is defined as
\begin{equation}
|{\hat \Omega}\ra = \frac{(u a^\dagger + v b^\dagger)^{2S}}{\sqrt{(2S)!}}|{\rm vac}\ra,
\end{equation}
where $(u, v)=(\cos(\theta/2)e^{i\phi/2}, \sin(\theta/2)e^{-i\phi/2})$ are spinor coordinates. Using $|{\hat \Omega}\ra$, the resolution of the identity is given by
\begin{equation}
I=\frac{2S+1}{4\pi}\int d{\hat \om} |{\hat \om}\ra\la{\hat \om}|,
\label{resol}
\end{equation}
where $I$ denotes a $(2S+1)$-dimensional identity matrix. Let us now consider the inner product  $\la k | l \ra$. We can set $k \ge l$ without loss of generality. 
Inserting the resolution of the identity (\ref{resol}) between $\la k|$ and $|l \ra$, the integral representation of the inner product can be obtained as
\begin{eqnarray}
\la k | l \ra &=& 
\prod_{j=1}^N(2S_j+1)!
 \int \prod_{j=1}^N \frac{d{\hat \Omega}_j}{4\pi} 
\Big(\frac{1-{\hat \om}_1\cdot{\hat \om}_{2}}{2}\Big)^{B_{12}-k}
\nonumber \\
&\times& \Big(\frac{1+{\hat \om}_1\cdot{\hat \om}_{2}}{2} -\cos\theta_1 \cos\theta_2 \Big)^{l}
K({\hat \om}_1, {\hat \om}_2)^{k-l} \nonumber \\
&\times& \prod_{i=2}^N \Big(\frac{1-{\hat \om}_i\cdot {\hat \om}_{i+1}}{2}\Big)^{B_{i,i+1}},
\label{Integral}
\end{eqnarray}
where 
\begin{equation}
K({\hat \om}_1, {\hat \om}_2)
=\frac{1}{2}(\cos\theta_1-\cos\theta_2 - i \sin\theta_1 \sin\theta_2 \sin(\phi_1-\phi_2)).
\nonumber
\end{equation}
Here we have already used the following relation: 
$\la {\rm vac}| a^{S-l}b^{S+l}|\om\ra = \sqrt{(2S)!}u^{S-l}v^{S+l}$.
First we consider the case where $k$ and $l$ have different parities. In this case, $k-l$ is odd and hence $K({\hat \om}_1, {\hat \om}_2)^{k-l}$ changes its sign under the change of variables $(\theta_j, \phi_j)$ to $(\pi-\theta_j, -\phi_j)$ ($j=1, 2, ..., N$). Since the other part of the integrand is invariant under this change of variables, we obtain $\la k|l \ra=0$. Therefore it is now obvious that ${\rm Re} \la k |l \ra =0$ when $k$ and $l$ have different parities.
Next we consider the case where $k$ and $l$ have the same parity. In this case, $k-l$ is even. Thus we set $k-l=2m$ ($m \in {\bf N}$) and expand $K({\hat \om}_1, {\hat \om}_2)^{2m}$ as
\begin{eqnarray}
K({\hat \om}_1, {\hat \om}_2)^{2m} &=&
 \Big( \frac{1}{2} \Big)^{2m}
\sum_{n=0}^{2m} {\binom{2m} n} (\cos\theta_1-\cos\theta_2)^{2m-n} \nonumber \\
&\times& (-i)^n (\sin\theta_1 \sin\theta_2 \sin (\phi_1-\phi_2))^n.
\end{eqnarray}
The imaginary part of  $K({\hat \om}_1, {\hat \om}_2)^{2m}$ comes from the contribution of the odd $n$'s in the above summation. Now we consider the following change of variables: $(\theta_j, \phi_j)$ to $(\theta_j, -\phi_j)$ ($j=1, 2, ..., N$). Under this transformation, ${\rm Im}[K({\hat \om}_1, {\hat \om}_2)^{2m}]$ changes its sign. On the other hand, the other part of the integrand in Eq. (\ref{Integral}) is real and invariant under this change of variables. Therefore, ${\rm Im} \la k| l \ra =0$ when $k$ and $l$ have the same parity. Finally, we remark that the generalization of the above result to the VBS state on an arbitrary graph is almost trivial since we have not used a specific property of the one-dimensional VBS state in our proof. 
\bibliography{manuscript}

\begin{thebibliography}{41}
\expandafter\ifx\csname natexlab\endcsname\relax\def\natexlab#1{#1}\fi
\expandafter\ifx\csname bibnamefont\endcsname\relax
  \def\bibnamefont#1{#1}\fi
\expandafter\ifx\csname bibfnamefont\endcsname\relax
  \def\bibfnamefont#1{#1}\fi
\expandafter\ifx\csname citenamefont\endcsname\relax
  \def\citenamefont#1{#1}\fi
\expandafter\ifx\csname url\endcsname\relax
  \def\url#1{\texttt{#1}}\fi
\expandafter\ifx\csname urlprefix\endcsname\relax\def\urlprefix{URL }\fi
\providecommand{\bibinfo}[2]{#2}
\providecommand{\eprint}[2][]{\url{#2}}

\bibitem[{\citenamefont{Hatsugai}(2006)}]{HatsugaiOrder3}
\bibinfo{author}{\bibfnamefont{Y.}~\bibnamefont{Hatsugai}},
  \bibinfo{journal}{J. Phys. Soc. Jpn.} \textbf{\bibinfo{volume}{75}},
  \bibinfo{pages}{123601} (\bibinfo{year}{2006}).

\bibitem[{\citenamefont{Wen}(1989)}]{TOrder}
\bibinfo{author}{\bibfnamefont{X.~G.} \bibnamefont{Wen}},
  \bibinfo{journal}{Phys. Rev. B} \textbf{\bibinfo{volume}{40}},
  \bibinfo{pages}{7387} (\bibinfo{year}{1989}).

\bibitem[{\citenamefont{Hatsugai}(2004)}]{HatsugaiOrder1}
\bibinfo{author}{\bibfnamefont{Y.}~\bibnamefont{Hatsugai}},
  \bibinfo{journal}{J. Phys. Soc. Jpn.} \textbf{\bibinfo{volume}{73}},
  \bibinfo{pages}{2604} (\bibinfo{year}{2004}).

\bibitem[{\citenamefont{Hatsugai}(2005)}]{HatsugaiOrder2}
\bibinfo{author}{\bibfnamefont{Y.}~\bibnamefont{Hatsugai}},
  \bibinfo{journal}{J. Phys. Soc. Jpn.} \textbf{\bibinfo{volume}{74}},
  \bibinfo{pages}{1374} (\bibinfo{year}{2005}).

\bibitem[{\citenamefont{Laughlin}(1981)}]{LaughlinEdge}
\bibinfo{author}{\bibfnamefont{R.~B.} \bibnamefont{Laughlin}},
  \bibinfo{journal}{Phys. Rev. B} \textbf{\bibinfo{volume}{23}},
  \bibinfo{pages}{5632} (\bibinfo{year}{1981}).

\bibitem[{\citenamefont{Hatsugai}(1993{\natexlab{a}})}]{HatsugaiEdge}
\bibinfo{author}{\bibfnamefont{Y.}~\bibnamefont{Hatsugai}},
  \bibinfo{journal}{Phys. Rev. Lett.} \textbf{\bibinfo{volume}{71}},
  \bibinfo{pages}{3697} (\bibinfo{year}{1993}{\natexlab{a}}).

\bibitem[{\citenamefont{Hatsugai}(1993{\natexlab{b}})}]{HatsugaiER}
\bibinfo{author}{\bibfnamefont{Y.}~\bibnamefont{Hatsugai}},
  \bibinfo{journal}{Phys. Rev. B} \textbf{\bibinfo{volume}{48}},
  \bibinfo{pages}{11851} (\bibinfo{year}{1993}{\natexlab{b}}).

\bibitem[{\citenamefont{Kennedy}(1990)}]{KennedyEdge}
\bibinfo{author}{\bibfnamefont{T.}~\bibnamefont{Kennedy}}, \bibinfo{journal}{J.
  Phys. Condens. Matter} \textbf{\bibinfo{volume}{2}}, \bibinfo{pages}{5737}
  (\bibinfo{year}{1990}).

\bibitem[{\citenamefont{Hagiwara et~al.}(1990)\citenamefont{Hagiwara,
  Katsumata, Affleck, Halperin, and Renard}}]{Hagiwara}
\bibinfo{author}{\bibfnamefont{M.}~\bibnamefont{Hagiwara}},
  \bibinfo{author}{\bibfnamefont{K.}~\bibnamefont{Katsumata}},
  \bibinfo{author}{\bibfnamefont{I.}~\bibnamefont{Affleck}},
  \bibinfo{author}{\bibfnamefont{B.~I.} \bibnamefont{Halperin}},
  \bibnamefont{and} \bibinfo{author}{\bibfnamefont{J.~P.}
  \bibnamefont{Renard}}, \bibinfo{journal}{Phys. Rev. Lett.}
  \textbf{\bibinfo{volume}{65}}, \bibinfo{pages}{3181} (\bibinfo{year}{1990}).

\bibitem[{\citenamefont{Su et~al.}(1979)\citenamefont{Su, Schrieffer, and
  Heeger}}]{SSH}
\bibinfo{author}{\bibfnamefont{W.~P.} \bibnamefont{Su}},
  \bibinfo{author}{\bibfnamefont{J.~R.} \bibnamefont{Schrieffer}},
  \bibnamefont{and} \bibinfo{author}{\bibfnamefont{A.~J.}
  \bibnamefont{Heeger}}, \bibinfo{journal}{Phys. Rev. Lett.}
  \textbf{\bibinfo{volume}{42}}, \bibinfo{pages}{1698} (\bibinfo{year}{1979}).

\bibitem[{\citenamefont{Read and Sachdev}(1989)}]{ReadSachdev}
\bibinfo{author}{\bibfnamefont{N.}~\bibnamefont{Read}} \bibnamefont{and}
  \bibinfo{author}{\bibfnamefont{S.}~\bibnamefont{Sachdev}},
  \bibinfo{journal}{Phys. Rev. Lett.} \textbf{\bibinfo{volume}{62}},
  \bibinfo{pages}{1694} (\bibinfo{year}{1989}).

\bibitem[{\citenamefont{Ryu and Hatsugai}(2006)}]{Ryu}
\bibinfo{author}{\bibfnamefont{S.}~\bibnamefont{Ryu}} \bibnamefont{and}
  \bibinfo{author}{\bibfnamefont{Y.}~\bibnamefont{Hatsugai}},
  \bibinfo{journal}{Phys. Rev. B} \textbf{\bibinfo{volume}{73}},
  \bibinfo{pages}{245115} (\bibinfo{year}{2006}).

\bibitem[{\citenamefont{Hirano and Hatsugai}(2007)}]{Hirano1}
\bibinfo{author}{\bibfnamefont{T.}~\bibnamefont{Hirano}} \bibnamefont{and}
  \bibinfo{author}{\bibfnamefont{Y.}~\bibnamefont{Hatsugai}},
  \bibinfo{journal}{J. Phys. Soc. Jpn.} \textbf{\bibinfo{volume}{76}},
  \bibinfo{pages}{074603} (\bibinfo{year}{2007}).

\bibitem[{\citenamefont{Katsura et~al.}(2007)\citenamefont{Katsura, Hirano, and
  Hatsugai}}]{Katsura}
\bibinfo{author}{\bibfnamefont{H.}~\bibnamefont{Katsura}},
  \bibinfo{author}{\bibfnamefont{T.}~\bibnamefont{Hirano}}, \bibnamefont{and}
  \bibinfo{author}{\bibfnamefont{Y.}~\bibnamefont{Hatsugai}},
  \bibinfo{journal}{Phys. Rev. B} \textbf{\bibinfo{volume}{76}},
  \bibinfo{pages}{012401} (\bibinfo{year}{2007}).

\bibitem[{\citenamefont{Kitaev and Preskill}(2006)}]{TEE1}
\bibinfo{author}{\bibfnamefont{A.}~\bibnamefont{Kitaev}} \bibnamefont{and}
  \bibinfo{author}{\bibfnamefont{J.}~\bibnamefont{Preskill}},
  \bibinfo{journal}{Phys. Rev. Lett.} \textbf{\bibinfo{volume}{96}},
  \bibinfo{pages}{110404} (\bibinfo{year}{2006}).

\bibitem[{\citenamefont{Levin and Wen}(2006)}]{TEE2}
\bibinfo{author}{\bibfnamefont{M.}~\bibnamefont{Levin}} \bibnamefont{and}
  \bibinfo{author}{\bibfnamefont{X.~G.} \bibnamefont{Wen}},
  \bibinfo{journal}{Phys. Rev. Lett.} \textbf{\bibinfo{volume}{96}},
  \bibinfo{pages}{110405} (\bibinfo{year}{2006}).

\bibitem[{\citenamefont{Berry}(1984)}]{Berryphase}
\bibinfo{author}{\bibfnamefont{M.~V.} \bibnamefont{Berry}},
  \bibinfo{journal}{Proc. R. Soc.} \textbf{\bibinfo{volume}{A392}},
  \bibinfo{pages}{45} (\bibinfo{year}{1984}).

\bibitem[{\citenamefont{Maruyama and Hatsugai}(2007)}]{Maruyama}
\bibinfo{author}{\bibfnamefont{I.}~\bibnamefont{Maruyama}} \bibnamefont{and}
  \bibinfo{author}{\bibfnamefont{Y.}~\bibnamefont{Hatsugai}},
  \bibinfo{journal}{J. Phys. Soc. Jpn.} \textbf{\bibinfo{volume}{76}},
  \bibinfo{pages}{113601} (\bibinfo{year}{2007}).

\bibitem[{\citenamefont{Hatsugai}(2007)}]{HatsugaiOrderP}
\bibinfo{author}{\bibfnamefont{Y.}~\bibnamefont{Hatsugai}},
  \bibinfo{journal}{J. Phys.: Condens. Matter} \textbf{\bibinfo{volume}{19}},
  \bibinfo{pages}{145209} (\bibinfo{year}{2007}).

\bibitem[{\citenamefont{Haldane}(1983{\natexlab{a}})}]{HaldaneGap}
\bibinfo{author}{\bibfnamefont{F.~D.~M.} \bibnamefont{Haldane}},
  \bibinfo{journal}{Phys. Lett. A} \textbf{\bibinfo{volume}{93}},
  \bibinfo{pages}{464} (\bibinfo{year}{1983}{\natexlab{a}}).

\bibitem[{\citenamefont{Ajiro et~al.}(1989)\citenamefont{Ajiro, Goto, Kikuchi,
  Sakakibara, and Inami}}]{NENP}
\bibinfo{author}{\bibfnamefont{Y.}~\bibnamefont{Ajiro}},
  \bibinfo{author}{\bibfnamefont{T.}~\bibnamefont{Goto}},
  \bibinfo{author}{\bibfnamefont{H.}~\bibnamefont{Kikuchi}},
  \bibinfo{author}{\bibfnamefont{T.}~\bibnamefont{Sakakibara}},
  \bibnamefont{and} \bibinfo{author}{\bibfnamefont{T.}~\bibnamefont{Inami}},
  \bibinfo{journal}{Phys. Rev. Lett.} \textbf{\bibinfo{volume}{63}},
  \bibinfo{pages}{1424} (\bibinfo{year}{1989}).

\bibitem[{\citenamefont{Granroth et~al.}(1996)\citenamefont{Granroth, Meisel,
  Chaparala, Jolicoeur, Ward, and Talham}}]{MnCl3}
\bibinfo{author}{\bibfnamefont{G.~E.} \bibnamefont{Granroth}},
  \bibinfo{author}{\bibfnamefont{M.~W.} \bibnamefont{Meisel}},
  \bibinfo{author}{\bibfnamefont{M.}~\bibnamefont{Chaparala}},
  \bibinfo{author}{\bibfnamefont{T.}~\bibnamefont{Jolicoeur}},
  \bibinfo{author}{\bibfnamefont{B.~H.} \bibnamefont{Ward}}, \bibnamefont{and}
  \bibinfo{author}{\bibfnamefont{D.~R.} \bibnamefont{Talham}},
  \bibinfo{journal}{Phys. Rev. Lett.} \textbf{\bibinfo{volume}{77}},
  \bibinfo{pages}{1616} (\bibinfo{year}{1996}).

\bibitem[{\citenamefont{Affleck}(1989)}]{thetaterm}
\bibinfo{author}{\bibfnamefont{I.}~\bibnamefont{Affleck}}, \bibinfo{journal}{J.
  Phys.: Condens. Matter} \textbf{\bibinfo{volume}{19}}, \bibinfo{pages}{3047}
  (\bibinfo{year}{1989}).

\bibitem[{\citenamefont{Kennedy and Tasaki}(1992{\natexlab{a}})}]{Z2Z2-1}
\bibinfo{author}{\bibfnamefont{T.}~\bibnamefont{Kennedy}} \bibnamefont{and}
  \bibinfo{author}{\bibfnamefont{H.}~\bibnamefont{Tasaki}},
  \bibinfo{journal}{Phys. Rev. B} \textbf{\bibinfo{volume}{45}},
  \bibinfo{pages}{304} (\bibinfo{year}{1992}{\natexlab{a}}).

\bibitem[{\citenamefont{Kennedy and Tasaki}(1992{\natexlab{b}})}]{Z2Z2-2}
\bibinfo{author}{\bibfnamefont{T.}~\bibnamefont{Kennedy}} \bibnamefont{and}
  \bibinfo{author}{\bibfnamefont{H.}~\bibnamefont{Tasaki}},
  \bibinfo{journal}{Commun. Math. Phys.} \textbf{\bibinfo{volume}{147}},
  \bibinfo{pages}{431} (\bibinfo{year}{1992}{\natexlab{b}}).

\bibitem[{\citenamefont{den Nijs and Rommelse}(1989)}]{Nijs}
\bibinfo{author}{\bibfnamefont{M.}~\bibnamefont{den Nijs}} \bibnamefont{and}
  \bibinfo{author}{\bibfnamefont{K.}~\bibnamefont{Rommelse}},
  \bibinfo{journal}{Phys. Rev. B} \textbf{\bibinfo{volume}{40}},
  \bibinfo{pages}{4709} (\bibinfo{year}{1989}).

\bibitem[{\citenamefont{Nakamura and Todo}(2002{\natexlab{a}})}]{Todo}
\bibinfo{author}{\bibfnamefont{M.}~\bibnamefont{Nakamura}} \bibnamefont{and}
  \bibinfo{author}{\bibfnamefont{S.}~\bibnamefont{Todo}}, \bibinfo{journal}{J.
  Phys. Soc. Jpn.} \textbf{\bibinfo{volume}{145}}, \bibinfo{pages}{217}
  (\bibinfo{year}{2002}{\natexlab{a}}).

\bibitem[{\citenamefont{Nakamura and Todo}(2002{\natexlab{b}})}]{Todo2}
\bibinfo{author}{\bibfnamefont{M.}~\bibnamefont{Nakamura}} \bibnamefont{and}
  \bibinfo{author}{\bibfnamefont{S.}~\bibnamefont{Todo}},
  \bibinfo{journal}{Phys. Rev. Lett.} \textbf{\bibinfo{volume}{89}},
  \bibinfo{pages}{077204} (\bibinfo{year}{2002}{\natexlab{b}}).

\bibitem[{\citenamefont{Hida}(1992)}]{Hida}
\bibinfo{author}{\bibfnamefont{K.}~\bibnamefont{Hida}}, \bibinfo{journal}{Phys.
  Rev. B} \textbf{\bibinfo{volume}{45}}, \bibinfo{pages}{2207}
  (\bibinfo{year}{1992}).

\bibitem[{\citenamefont{Yamanaka et~al.}(1996)\citenamefont{Yamanaka, Oshikawa,
  and Miyashita}}]{OshikawaDimer}
\bibinfo{author}{\bibfnamefont{M.}~\bibnamefont{Yamanaka}},
  \bibinfo{author}{\bibfnamefont{M.}~\bibnamefont{Oshikawa}}, \bibnamefont{and}
  \bibinfo{author}{\bibfnamefont{S.}~\bibnamefont{Miyashita}},
  \bibinfo{journal}{J. Phys. Soc. Jpn.} \textbf{\bibinfo{volume}{65}},
  \bibinfo{pages}{1562} (\bibinfo{year}{1996}).

\bibitem[{\citenamefont{Haldane}(1983{\natexlab{b}})}]{HaldaneLargeD}
\bibinfo{author}{\bibfnamefont{F.~D.~M.} \bibnamefont{Haldane}},
  \bibinfo{journal}{Phys. Rev. Lett.} \textbf{\bibinfo{volume}{50}},
  \bibinfo{pages}{1153} (\bibinfo{year}{1983}{\natexlab{b}}).

\bibitem[{\citenamefont{Aschauer and Schollw{\"o}ck}(1998)}]{SchollwockDterm}
\bibinfo{author}{\bibfnamefont{H.}~\bibnamefont{Aschauer}} \bibnamefont{and}
  \bibinfo{author}{\bibfnamefont{U.}~\bibnamefont{Schollw{\"o}ck}},
  \bibinfo{journal}{Phys. Rev. B} \textbf{\bibinfo{volume}{58}},
  \bibinfo{pages}{359} (\bibinfo{year}{1998}).

\bibitem[{\citenamefont{Oshikawa}(1992)}]{OshikawaDterm}
\bibinfo{author}{\bibfnamefont{M.}~\bibnamefont{Oshikawa}},
  \bibinfo{journal}{J. Phys.: Condens. Matter} \textbf{\bibinfo{volume}{4}},
  \bibinfo{pages}{7469} (\bibinfo{year}{1992}).

\bibitem[{\citenamefont{Oshikawa et~al.}(1995)\citenamefont{Oshikawa, Yamanaka,
  and Miyashita}}]{OshikawaDterm2}
\bibinfo{author}{\bibfnamefont{M.}~\bibnamefont{Oshikawa}},
  \bibinfo{author}{\bibfnamefont{M.}~\bibnamefont{Yamanaka}}, \bibnamefont{and}
  \bibinfo{author}{\bibfnamefont{S.}~\bibnamefont{Miyashita}},
  \bibinfo{journal}{condmat/9507098 (unpublished)}  (\bibinfo{year}{1995}).

\bibitem[{\citenamefont{Fukui et~al.}(2006)\citenamefont{Fukui, Hatsugai, and
  Suzuki}}]{Fukui}
\bibinfo{author}{\bibfnamefont{T.}~\bibnamefont{Fukui}},
  \bibinfo{author}{\bibfnamefont{Y.}~\bibnamefont{Hatsugai}}, \bibnamefont{and}
  \bibinfo{author}{\bibfnamefont{H.}~\bibnamefont{Suzuki}},
  \bibinfo{journal}{J. Phys. Soc. Jpn.} \textbf{\bibinfo{volume}{74}},
  \bibinfo{pages}{1674} (\bibinfo{year}{2006}).

\bibitem[{\citenamefont{Kitazawa and Nomura}(1997)}]{Kitazawa}
\bibinfo{author}{\bibfnamefont{A.}~\bibnamefont{Kitazawa}} \bibnamefont{and}
  \bibinfo{author}{\bibfnamefont{K.}~\bibnamefont{Nomura}},
  \bibinfo{journal}{J. Phys. Soc. Jpn.} \textbf{\bibinfo{volume}{66}},
  \bibinfo{pages}{3379} (\bibinfo{year}{1997}).

\bibitem[{\citenamefont{Affleck et~al.}(1987)\citenamefont{Affleck, Kennedy,
  Lieb, and Tasaki}}]{AKLT}
\bibinfo{author}{\bibfnamefont{I.}~\bibnamefont{Affleck}},
  \bibinfo{author}{\bibfnamefont{T.}~\bibnamefont{Kennedy}},
  \bibinfo{author}{\bibfnamefont{E.~H.} \bibnamefont{Lieb}}, \bibnamefont{and}
  \bibinfo{author}{\bibfnamefont{H.}~\bibnamefont{Tasaki}},
  \bibinfo{journal}{Phys. Rev. Lett.} \textbf{\bibinfo{volume}{59}},
  \bibinfo{pages}{799} (\bibinfo{year}{1987}).

\bibitem[{\citenamefont{Dillenschneider
  et~al.}(2007)\citenamefont{Dillenschneider, Kim, and Han}}]{chiralVBS}
\bibinfo{author}{\bibfnamefont{R.}~\bibnamefont{Dillenschneider}},
  \bibinfo{author}{\bibfnamefont{J.~H.} \bibnamefont{Kim}}, \bibnamefont{and}
  \bibinfo{author}{\bibfnamefont{J.~H.} \bibnamefont{Han}},
  \bibinfo{journal}{cond-mat/0705.3993}  (\bibinfo{year}{2007}).

\bibitem[{\citenamefont{Auerbach}(1998)}]{Auelbach}
\bibinfo{author}{\bibfnamefont{A.}~\bibnamefont{Auerbach}},
  \emph{\bibinfo{title}{Interacting Electrons and Quantum Magnetism}}
  (\bibinfo{publisher}{Springer, New York}, \bibinfo{year}{1998}).

\bibitem[{\citenamefont{Kirillov and Korepin}(1990)}]{arbitVBS}
\bibinfo{author}{\bibfnamefont{A.~N.} \bibnamefont{Kirillov}} \bibnamefont{and}
  \bibinfo{author}{\bibfnamefont{V.~E.} \bibnamefont{Korepin}},
  \bibinfo{journal}{Sankt Petersburg Mathematical Journal}
  \textbf{\bibinfo{volume}{1}}, \bibinfo{pages}{47} (\bibinfo{year}{1990}).

\bibitem[{\citenamefont{Fan et~al.}(2004)\citenamefont{Fan, Korepin, and
  Roychowdhury}}]{Fan1}
\bibinfo{author}{\bibfnamefont{H.}~\bibnamefont{Fan}},
  \bibinfo{author}{\bibfnamefont{V.}~\bibnamefont{Korepin}}, \bibnamefont{and}
  \bibinfo{author}{\bibfnamefont{V.}~\bibnamefont{Roychowdhury}},
  \bibinfo{journal}{Phys. Rev. Lett.} \textbf{\bibinfo{volume}{93}},
  \bibinfo{pages}{227203} (\bibinfo{year}{2004}).

\end{thebibliography}
\end{appendix}
\end{document}